# Thermodynamic behavior of correlated electron-hole fluids in van der Waals heterostructures

Ruishi Qi[1,2,†], Andrew Y. Joe[1,2,†,*], Zuocheng Zhang[1], Yongxin Zeng[3], Tiancheng Zheng[1,4], Qixin Feng[1,2], Emma Regan[1,2,5], Jingxu Xie[2,5], Zheyu Lu[2,5], Takashi Taniguchi[6], Kenji Watanabe[7], Sefaattin Tongay[8], Michael F. Crommie[1,2], Allan H. MacDonald[3], Feng Wang[1,2,9,*]

[1] Department of Physics, University of California, Berkeley, CA 94720, USA

[2] Materials Sciences Division, Lawrence Berkeley National Laboratory, Berkeley, CA 94720, USA

[3] Department of Physics, University of Texas at Austin, Austin, Texas 78712, USA

[4] School of Physical Sciences, University of Chinese Academy of Sciences, Beijing, China.

[5] Graduate Group in Applied Science and Technology, University of California at Berkeley, Berkeley, CA 94720, USA

[6] International Center for Materials Nanoarchitectonics, National Institute for Materials Science, 1-1 Namiki, Tsukuba 305-0044, Japan

[7] Research Center for Functional Materials, National Institute for Materials Science, 1-1 Namiki, Tsukuba 305-0044, Japan

[8] School for Engineering of Matter, Transport and Energy, Arizona State University, Tempe, AZ 85287, USA

[9] Kavli Energy NanoSciences Institute, University of California Berkeley and Lawrence Berkeley National Laboratory, Berkeley, CA 94720, USA.

† These authors contributed equally.

* To whom correspondence should be addressed: andrew.joe@ucr.edu, fengwang76@berkeley.edu

**Coupled two-dimensional electron-hole bilayers provide a unique platform to study strongly correlated Bose-Fermi mixtures in condensed matter. Electrons and holes in spatially separated layers can bind to form interlayer excitons, composite Bosons expected to support high-temperature exciton superfluids. The interlayer excitons can also interact strongly with excess charge carriers when electron and hole densities are unequal. Here, we use optical spectroscopy to quantitatively probe the local thermodynamic properties of strongly correlated electron-hole fluids in $MoSe_2/hBN/WSe_2$ heterostructures. We observe a discontinuity in the electron and hole chemical potentials at matched electron and hole densities, a definitive signature of an excitonic insulator ground state. The excitonic insulator is stable up to a Mott density of ~$0.8 \times 10^{12}$ cm$^{-2}$ and has a thermal ionization temperature of ~70 K. The density dependence of the electron, hole, and exciton chemical potentials reveals strong correlation effects across the phase diagram. Compared with a non-interacting uniform charge distribution, the correlation effects lead to significant attractive exciton-exciton and exciton-charge interactions in the electron-hole fluid. Our work highlights the unique quantum behavior that can emerge in strongly correlated electron-hole systems.**

Two-dimensional (2D) electron gases have been extensively studied in the last few decades, leading to fascinating discoveries such as the integer and fractional quantum Hall effects and 2D Wigner crystals[1–5]. Equilibrium 2D electron and hole gases that are in close proximity to each other while remaining electrically isolated[6] have been difficult to realize experimentally and are therefore much less explored. The strong attractive interactions between electrons and holes are expected to play a significant role in the thermodynamic behavior of the coupled system. In the limit of low and matched electron and hole densities, the layer-separated electrons and holes can form bound pairs to create interlayer excitons. Interlayer excitons are composite bosons and are therefore an attractive candidate for studying correlated quantum Boson states in condensed matter[7–10], whose phase diagrams are expected to include exciton Bose-Einstein condensates (BECs)[11–13], excitonic insulators[14,15], and exciton crystals[16]. In the general case of unequal electron and hole densities, the system provides a unique platform to study a tunable mixture of equilibrium Fermi and Bose gases that contain charge-exciton complexes.

Semiconducting transition metal dichalcogenide (TMD) heterostructures, due to their strong Coulomb interaction and large exciton binding energies[17–19], provide a model system to explore correlated 2D electron-hole (e-h) fluids. By inserting a thin tunnel barrier such as hexagonal boron nitride (hBN) between two TMD layers, ultralong exciton lifetimes, high exciton density, and equilibrium e-h and exciton fluids can be realized[14,20]. Theoretical studies of such heterostructures have predicted record-high BEC critical temperatures with a unique tunability to explore strong and weak coupling regimes[6,12,21,22]. Previously, electrical capacitance measurements have been employed to probe interlayer excitons in $MoSe_2$/hBN/$WSe_2$ heterostructures and revealed the presence of correlated excitonic insulators[14], defined as a charge-incompressible but exciton-compressible state. However, thermodynamic behavior of the e-h fluid across the electron and hole doped phase diagram has not been established.

Here, we develop quantitative optical spectroscopy to accurately determine the local electron and hole density and use it to determine the thermodynamic behavior of correlated e-h fluids in TMD heterostructures. It is well known that optical resonances in TMD layers depend sensitively on the carrier density due to their enhanced light-matter and Coulomb interactions[23,24]. With careful in-situ calibration of this dependence, we can determine the electron, hole, and exciton densities, as a function of the electron, hole, and exciton chemical potentials of our gated TMD heterostructures. We observe electrically tunable and highly correlated excitons and exciton-charge complexes in this system: (1) An excitonic insulator ground state can appear at equal electron and hole density, which features strong exciton-exciton correlation. The exciton insulator ionizes at both increased exciton density (Mott transition) and elevated temperature (thermal melting). (2) The interlayer excitons interact strongly with additional electrons or holes, leading to an exciton-density dependent charge chemical potential for a fixed net charge density.

Figure 1a shows a schematic cross-section and electrical circuit representation of our device. We use $MoSe_2$ and $WSe_2$ as the active electron and hole layers and separate them by a 3 nm thin hBN layer. The heterostructure is encapsulated with dielectric hBN and graphite top ($V_{TG}$) and bottom ($V_{BG}$) gates. To create reliable electrical contacts, we follow Ref. 14 to incorporate a thicker hBN spacer between the TMD layers in part of the heterostructure (region 0). The TMD layers are then contacted by few-layer graphene in region 0. The vertical electric field generated by $V_\Delta \equiv V_{TG} - V_{BG}$ can control the electric potential difference between the $MoSe_2$ and $WSe_2$ layers in region 0. It enables a heavily electron doped $MoSe_2$ layer and a heavily hole doped $WSe_2$ layer in region 0, which serve as carrier reservoirs for the area of interest (region 1) (details in Methods and Extended Data Figs. 1-2). This improved contact method allows the simultaneous doping of both electrons and holes in the active region. The combination of the thin hBN layer suppressing interlayer tunneling and efficient charge injection from the contacts ensures an equilibrium state and accurate measurements of thermodynamic quantities. An optical image of such a device is shown in Figure 1b.

We control the band alignment and electron ($n_e$) and hole ($n_h$) densities in the active region using gate voltages ($V_{TG}$ and $V_{BG}$) and the WSe$_2$ voltage ($V_h$) while grounding the MoSe$_2$ layer ($V_e = 0$). The interlayer bias voltage $V_B \equiv V_h - V_e$ and vertical electric field $V_\Delta$ both create an energy difference in the two layers and tune the effective band gap. The gate voltage $V_G \equiv V_{TG} + V_{BG}$ tunes the Fermi level and controls the net charge $n_e - n_h$ in the e-h system. We can achieve reliable electrical control in multiple devices and the data shown below is highly reproducible (Extended Data Figs. 1, 3 and 4). Unless otherwise specified, the results below are from device D1 at $V_\Delta = 11$ V at sample temperature $T = 2.5$ K.

We use optical spectroscopy to quantitatively determine the electron and hole densities in MoSe$_2$ and WSe$_2$ layers, respectively. In Figure 1c, we first show representative reflection spectra $R$ as a function of the applied gate voltage $V_G$ in the low $V_B$ regime where the type-II band gap is not closed. When the Fermi level is tuned into the gap and both layers are undoped (middle region of Fig. 1c), we observe strong resonances at 719 nm and 754 nm, which are attributed to the intralayer excitonic responses in the WSe$_2$ and MoSe$_2$, respectively. When the TMD is doped with holes or electrons (left and right regions), three-particle bound states commonly known as trions will form, giving rise to an additional absorption peak at longer wavelength[23,25]. With increasing doping, the trion peak becomes stronger, while the exciton peak blueshifts and loses its oscillator strength due to Pauli blocking[24]. Thus, the reflection spectrum provides a sensitive probe for the carrier density in each layer. In the well-understood single layer doped regime, the undoped TMD layer simply acts as a passive dielectric layer, and the doping density in the active TMD layer can be calculated from a simple parallel capacitor model (details in Methods). This allows us to create a map between the reflection spectra and carrier density for each layer (Extended Data Fig. 5 and faded colored lines in Fig. 1d).

When the bandgap is closed by a large $V_B$, both layers become doped, forming a correlated e-h fluid. The black dots in Fig. 1d show a representative spectrum, where trion peaks in both layers appear in this regime. The faded lines (Fig. 1d) show the single-layer spectra for electron and hole densities from 0 - 2 × 10$^{12}$ cm$^{-2}$ taken from the spectrum-density maps. We match the measured reflection spectrum with the single-layer spectrum-density maps, allowing us to extract the electron and hole densities in this less-understood e-h fluid regime. The solid lines display the matched single-layer spectra for the representative spectrum, which give $n_e = 1.11 \times 10^{12}$ cm$^{-2}$ and $n_h = 0.71 \times 10^{12}$ cm$^{-2}$. This spectrum-matching method assumes that the intralayer exciton/trion states do not get significantly perturbed by the existence of charges in the other TMD layer. The intralayer excitons and trions are both tightly bound particles with a typical size of only ~1 nm[26–29], and their optical transitions are only minimally affected by charged particles in the other TMD layer that is 3 nm away. In our experiment, the excellent match in wavelength, intensity and line shape for all peaks confirms that our assumption is valid.

Figs. 2a-b show the gate and bias voltage dependence of the measured electron and hole densities. From the density maps, we can establish the phase diagram of the carrier doping in the MoSe$_2$/hBN/WSe$_2$ heterostructure (Fig. 2c). At low $V_B$, the type-II band-aligned heterostructure has a finite band gap, and the gate voltage can tune the system from charge neutrality (black) to hole-doped (green) or electron doped (red). With increasing bias voltage, the band gap decreases until it closes at a critical $V_B \approx 0.82$ V. Further increasing $V_B$ dopes both layers simultaneously

(yellow). This is where interlayer exciton formation is expected. Notably, the carrier density contour lines (dash-dotted lines in Figs. 2a-b) are not straight in this regime. Instead, they show kinks near $n_e \approx n_h$, indicating that the e-h fluid is strongly correlated.

The voltage dependence of $n_e$ and $n_h$ allows us to determine the compressibility of interlayer excitons and charge carriers, independently. Figure 2d shows the exciton compressibility obtained by plotting the partial derivative of exciton density $n_x = \min(n_e, n_h)$ with respect to $V_B$. Figure 2e shows the charge compressibility, defined as the partial derivative of the net charge $n_e - n_h$ with respect to $V_G$. Interestingly, we find there is a triangle region that extends beyond the exciton doping edge (black dashed line) where there is a charge-incompressible, exciton-compressible state. Previous capacitance studies have shown similar results and have identified this state as an excitonic insulator[14].

Another important consequence of the formation of interlayer excitons is the scaling of the carrier density as a function of the bias voltage at net charge neutrality. In Figure 2f, we show the bias dependence of $n_x$ along the $n_e \approx n_h$ line (white dashed line in Figs. 2a-b). Interestingly, we observe a sublinear increase of the exciton density. Fitting the experimental data to a power law scaling $n_x \propto \Delta V_B^\eta$ gives a fitted power of $\eta \approx 0.76 \pm 0.02$ (Fig. 2f inset). This scaling power is consistent with previous theoretical studies of strongly correlated dipolar exciton BECs[11], where quantum Monte Carlo simulations predict $\eta \approx 0.74$.

In the non-interacting limit, the electron and hole densities should increase linearly with $V_B$ as determined by the geometric capacitance (yellow dashed line). The deviation from the linear bias dependence arises from the exciton-exciton interactions. Such exciton-exciton interactions can be understood in two different physical pictures. In physical picture I, we can consider the net repulsive dipolar interaction between tightly bound excitons. This repulsion between dipoles causes an energy penalty to generate a high-density exciton fluid. It therefore leads to a sublinear scaling where the compressibility at low exciton density is higher than that at high exciton density. In physical picture II, we consider the exciton compressibility with respect to the "ideal capacitor" model, which assumes homogeneous electron-hole distributions. We find that the exciton compressibility approaches the value predicted by the geometric capacitance at high density, but has a significantly enhanced value at low exciton density. It suggests a strongly attractive exciton-exciton interaction at low exciton density if we use the homogeneous electron-hole distribution as the reference point. In this picture, we can focus on the pure correlation effect by separating out the electrostatic energy associated with uniform electron and hole distributions. The second picture is widely used in the study of correlation effects in electron liquids[30–32], and it is the only picture that can be readily used to describe the general electron-hole liquid that includes the electron-hole plasma at high density as well as unbalanced electron and hole densities.

To further quantify the thermodynamic behavior of the interlayer excitons and the exciton-charge complexes in the e-h fluid, we adopt the second picture to determine the electron, hole, and exciton chemical potential as a function of the electron and hole densities. Following the theory developed in Ref. 6, we separate the electrochemical potential into the electric potential $\phi$, which contains the electrostatic energy associated with homogeneous charge distributions in an ideal capacitor and is device-geometry dependent, and the chemical potential $\mu$, which contains the correlation effects in the e-h fluids. At equilibrium the chemical potentials $(\mu_e, \mu_h)$ of the electron layer and the hole layer are related to the electric potentials $(\phi_e, \phi_h)$ by

$$-e\phi_e + \mu_e = -eV_e - E_c,$$
$$e\phi_h + \mu_h = eV_h - E_v, \qquad (1)$$

where $e$ is elementary charge, and $E_c$ and $E_v$ are the conduction and valence band energies. From the experimentally determined carrier densities and applied voltages, the electric potentials can be derived from the Poisson equation

$$\frac{-en_e}{\epsilon_0 \epsilon_{BN}} = \frac{\phi_e - V_{TG}}{t_t} + \frac{\phi_e - \phi_h}{t_m},$$
$$\frac{en_h}{\epsilon_0 \epsilon_{BN}} = \frac{\phi_h - V_{BG}}{t_b} + \frac{\phi_h - \phi_e}{t_m}. \qquad (2)$$

Here $\epsilon_0$ is vacuum permittivity, $\epsilon_{BN} = 4.2$ is the out-of-plane dielectric constant for hBN[33], and $t_t, t_m, t_b$ are the top, middle, and bottom hBN thicknesses, respectively. The chemical potentials can then be experimentally determined as a function of $n_e$ and $n_h$, as shown in Figures 3a-b.

Notably, the electron chemical potential (Fig. 3a) has a discontinuity across the diagonal line ($n_e = n_h$), which is more prominent at low densities, and disappears above a critical density. The hole chemical potential also shows a similar jump but with opposite sign (Fig. 3b). In Figure 3c, we show a typical chemical potential linecut at fixed hole doping $n_h = 0.15 \times 10^{12}$ cm$^{-2}$. By fitting it to a linear background plus a step function, we estimate the chemical potential jump ($\Delta\mu_e, \Delta\mu_h$) at this density to be +13 meV and -13 meV for the electrons and holes, respectively. This chemical potential discontinuity effectively opens a gap at $n_e = n_h$ for single charge injection, leading to a charge-incompressible phase at finite density. Nevertheless, the exciton chemical potential $\mu_x = \mu_e + \mu_h$ is continuous (Fig. 3d), and therefore excitons are compressible.

The chemical potential jump is another definitive signature of interlayer excitons in the excitonic insulator phase. Close to the excitonic insulator phase, the system consists of $n_x = \min(n_e, n_h)$ interlayer excitons and $|n_e - n_h|$ unpaired charges. When $n_e < n_h$, adding one more electron to the system creates one interlayer exciton and removes one free hole, gaining an exciton binding energy. When $n_e > n_h$, however, adding one electron does not change the number of excitons. Thus, in the low-density limit the chemical potential jump at $n_e = n_h$ gives the interlayer exciton binding energy $\epsilon_b$.

At higher densities, quantum dissociation of the interlayer exciton is expected after the Mott limit is reached[34]. Figure 3e shows the density dependence of the measured chemical potential jump. The binding energy in the dilute limit, $\epsilon_b \approx 20$ meV, is consistent with previous theoretical[21] and experimental studies[14]. The chemical potential jump decreases with density and vanishes above the Mott density ~ $0.8 \times 10^{12}$ cm$^{-2}$, above which the excitons dissociate, and the system turns into a degenerate electron-hole Fermi gas[21].

We next examine the general thermodynamic behavior of the correlated e-h fluid across the phase diagram. To provide an approximate framework to understand the behavior, we can parameterize the total energy of the e-h fluid as

$$E(n_e, n_h) = -\epsilon_b n_x + \frac{\alpha}{2}(n_e - n_h)^2 + \frac{g}{2}n_x^2 + \beta n_x |n_e - n_h|. \qquad (3)$$

Here $\alpha$, $g$, and $\beta$ characterize the charge-charge interaction strength, exciton-exciton interaction strength, and exciton-charge interaction strength, respectively. The electron, hole, and exciton chemical potentials are given by partial derivatives of $E$ with respect to $n_e$, $n_h$ and $n_x$ respectively, which yield

$$\mu_e = \frac{\partial E}{\partial n_e} = \begin{cases} \alpha(n_e - n_h) + \beta n_h, & (n_e > n_h) \\ \alpha(n_e - n_h) + \beta n_h - \epsilon_b + (g - 2\beta)n_e, & (n_e < n_h) \end{cases}, \quad (4)$$

$$\mu_h = \frac{\partial E}{\partial n_h} = \begin{cases} \alpha(n_h - n_e) + \beta n_e, & (n_h > n_e) \\ \alpha(n_h - n_e) + \beta n_e - \epsilon_b + (g - 2\beta)n_h, & (n_h < n_e) \end{cases}, \quad (5)$$

$$\mu_x = \mu_e + \mu_h = -\epsilon_b + g n_x + \beta |n_e - n_h|. \quad (6)$$

Our approximate energy expression gives an empirical linear expansion of the chemical potentials in the low doping regime. In Fig. 3f, we show a linecut of $\mu_e$ and $\mu_h$ at zero-hole density (corresponding to a linecut along the vertical axis in Fig. 3a-b). The electron chemical potential (red circles in Fig. 3f) decreases strongly with $n_e$. This behavior is controlled by the charge-charge interaction $\alpha$. Fitting the experimental linecut to a polynomial and comparing it to the ansatz electron chemical potential $\mu_e(n_e, n_h = 0) = \alpha n_e$ gives an initial slope of $\alpha = -66 \pm 6$ meV/($10^{12}$cm$^{-2}$). The negative $\alpha$ is a manifestation of the negative compressibility in low-density electron gases and arises from strong exchange interactions[30].

In contrast, the hole chemical potential $\mu_h$ increases significantly with the electron density at zero-hole density (green circles in Fig. 3f). This unusual behavior arises from a competition between the charge-charge interaction and the charge-exciton interaction, and it highlights the strong interlayer correlation effects. With the interlayer interactions, the addition of the hole can be viewed as simultaneously adding an exciton and removing an electron. It results in the hole chemical potential described by $\mu_h(n_e, n_h = 0) = -\epsilon_b + (\beta - \alpha)n_e$. The positive slope in Fig. 3f is determined by $\beta - \alpha$, with $\beta = -35 \pm 6$ meV/($10^{12}$cm$^{-2}$).

Next, we examine the evolution of the exciton chemical potential $\mu_x$. Figure 3g shows that the exciton chemical potential decreases with the exciton density at $n_e = n_h$ (corresponding to a diagonal linecut of Fig. 3d). Under this condition, the exciton chemical potential $\mu_x = -\epsilon_b + g n_x$. Figure 3g shows that the effective interaction between excitons is strongly attractive with $g = -70 \pm 2$ meV/($10^{12}$cm$^{-2}$). A negative $g$ is consistent with the physical picture II described above, where the attractive intralayer exchange interaction, in reference to a homogeneous charge distribution, dominates over the repulsive interlayer exchange interaction[12,6]. At higher doping the chemical potential deviates from a linear line, suggesting a positive second order interaction $19 \pm 2$ meV/($10^{12}$cm$^{-2}$)$^2$ as predicted in previous theoretical studies[6].

Figure 3h further displays the evolution of the exciton chemical potential with the net charge density $n_e - n_h$ at fixed exciton density $n_x = 0.05 \times 10^{12}$ cm$^{-2}$. $\mu_x$ has a maximum value at the charge neutral state, and it decreases with either additional electron or hole doping, indicating an attractive interaction between the exciton and charge in the exciton-charge complex. Using the expression $\mu_x = -\epsilon_b + g n_x + \beta |n_e - n_h|$, we can deduce an attractive exciton-charge interaction strength $\beta = -39 \pm 2$ meV/($10^{12}$cm$^{-2}$). This value agrees with that value obtained from Fig. 3f.

Although we have focused on only a few special line cuts in Fig. 3f-h, the parameterized chemical potential (Eqs. 4-6) provides a good description of the complete 2D phase diagram of

Fig. 3a-b at low exciton and charge density regime. Furthermore, our experimental chemical potential maps can be reproduced semi-quantitatively by our mean-field theory calculations (Methods and Extended Data Fig. 6). The mean-field calculations clearly predict the chemical potential discontinuity at $n_e = n_h$. They also reproduce the attractive charge-charge, charge-exciton, and exciton-exciton interactions. However, the mean-field theory overestimates the exciton binding energy and cannot capture the Mott transition due to the lack of screening in the calculation.

Lastly, we investigate the temperature dependence of the chemical potential in the e-h fluid. Figure 4a shows the chemical potential jump at neutrality ($\Delta\mu = (\Delta\mu_e - \Delta\mu_h)/2$) as a function of density and temperature. With increasing temperature, the magnitude of the discontinuity decreases, suggesting ionization of the interlayer excitons. At the lowest density, we estimate the interlayer excitons ionize around ~70 K. Similarly, the Mott density decreases with increasing temperature, which can be qualitatively understood as temperature aiding the exciton ionization. Figure 4b shows the exciton chemical potential change as a function of the exciton density along the $n_e = n_h$ condition (similar to Fig. 3g) for various temperatures. We find that the exciton-exciton interaction strength remains constant until ~40 K and then becomes weaker with increasing temperature (Fig. 4b inset). In Fig. 4c, we show linecuts of $\mu_x$ at constant $n_x$ (similar to Fig. 3h) for various temperatures. The cusp at charge neutrality becomes broader with increasing temperature, indicating reduced exciton-charge interaction strength. The fitted $\beta$ also remains constant up to ~40 K and then weakens (Fig. 4c inset). The change in $g$ and $\beta$ occurs before the disappearance of the chemical potential jump, implying a change in the interaction behavior of the exciton-charge complex before the complete thermal ionization of the interlayer excitons.

Our work opens many exciting directions for future studies of interlayer excitons and the search for the exciton condensate at higher temperatures without magnetic fields. The measurements of the chemical potential and interaction strengths allow direct connection to several theoretical predictions[6,11,12,21] and serve as a guide for future studies on the interactions in the electron-hole fluids and possible exciton condensates.

## Methods
**Device fabrication.** We use a dry-transfer method based on polyethylene terephthalate glycol (PETG) stamps to fabricate the heterostructures. Monolayer MoSe$_2$, monolayer WSe$_2$, few-layer graphene and hBN flakes are mechanically exfoliated from bulk crystals onto Si substrates with a 90-nm-thick SiO$_2$ layer. We use 15-25 nm hBN as the gate dielectric, 20-30 nm hBN for the interlayer spacer in region 0, and 2-3 nm thin hBN as the interlayer spacer in region 1. A 0.5 mm thick clear PETG stamp is employed to sequentially pick up the flakes at 65-75 °C. The whole stack is then released onto a 90 nm SiO$_2$/Si substrate at 95-100 °C, followed by dissolving the PETG in chloroform at room temperature for at least one day. Electrodes (50 nm Au with 5 nm Cr adhesion layer) are defined using photolithography (Durham Magneto Optics, MicroWriter) and electron beam evaporation (Angstrom Engineering).

Extended Data Fig. 1 shows detailed device structures and hBN thickness values for three devices D1 - D3. In addition to the structure described above, D1 has another region (region 2) that is controlled by another back gate with a thicker dielectric hBN. Data taken in region 2 of D1, D2 and D3 are consistent with the main text (Extended Data Figs. 3-4).

**Electrical contacts and measurements.** Keithley 2400 or 2450 source meters are used for applying gate and bias voltages and monitoring the leakage current. In all measurements, the leakage current is kept below 5 pA to make sure the system is in an equilibrium state. The interlayer leakage current gives a lower bound of the exciton recombination lifetime $\tau$. We assume all the interlayer leakage is from electron-hole recombination, and the lifetime can be estimated as $\tau = en_x A/I$, where $A \approx 136$ μm² is the region 1 heterostructure area and $I$ is the leakage current. As shown in Extended Data Fig. 2, the average lifetime ~320 ms is orders of magnitude longer than charge injection time from the contact, which is on the millisecond time scale.

To make good contacts to the TMD layers and achieve equilibrium states, a thick hBN layer (20-30 nm) is inserted between the TMD layers in the contact region (region 0). The vertical electric field $V_\Delta$ creates a much larger voltage difference in region 0 than in region 1, so the band gap in region 0 will be easily closed. Thus, for the voltage range we are interested in, region 0 sustains high doping in both layers and forms better electrical contact to the graphite electrodes. Extended Data Fig. 2c-d shows the doping phase diagram for region 0 and region 1 at constant bias volage $V_B = 1$ V. The scan is performed in a "snake" order, i.e., columns have alternating scan directions. We observe hysteresis in region 0 doping, suggesting that the contact is not effective and has a long response time at low doping levels. In region 1, when region 0 is not heavily doped, we observe hysteresis; but when region 0 has strong electron and hole doping (dashed box), the contact resistance reduces significantly, resulting in a clean phase diagram with no hysteresis. The data shown in the main text is at $V_\Delta = 11$ V, where region 0 is at a very high doping level. It is also taken as a "snake" scan, and we do not observe any hysteresis down to mV level, indicating our system is in equilibrium with the contact.

**Calibration of hBN thickness.** The dielectric hBN (~ 20 nm) thickness is determined by optical contrast calibrated by atomic force microscopy (AFM). We perform AFM thickness measurement and optical microscope imaging for tens of hBN flakes ranging from 1 to 70nm. The AFM is operated in contact mode to get accurate thickness values. The optical microscope images are taken with a colored camera with fixed color temperature and ISO settings. The RGB optical contrast of the hBN flake, defined as (hBN – substrate)/substrate for the red, green, and blue (RGB) channels of the image, is shown in Extended Data Fig. 7a. We fit the RGB contrast to a 4[th] order polynomial. The measured data points and the fitted curve agree nicely with theoretically calculated reflectivity using Fresnel equations (Extended Data Fig. 7b). Then the thickness of a given hBN flake can be determined by matching its optical contrast with the fitted curve. For ~ 20 nm thick hBN flakes, the accuracy is typically within 1 nm. For the thin hBN spacers, the relative error of this method will be worse, so we accurately determine their thickness from reflection measurements of the fabricated heterostructure. We keep the system net charge neutral while scanning the $V_B$ and $V_\Delta$. The exciton gap is closed when

$$E_g - eV_B - \frac{et_m}{t_t + t_m + t_b} V_\Delta = 0, \tag{7}$$

where $E_g$ is the band gap energy. The slope of the gap-closing boundary gives the ratio of interlayer distance to total hBN thickness. Extended Data Fig. 7c shows a line fit of the gap-closing boundary for D1, which gives $t_m = 3.0 \pm 0.2$ nm (interpreted as the effective interlayer distance, which is slightly larger than the thin hBN thickness due to finite TMD layer thickness), consistent with the value determined from optical contrast (2.7 ± 0.6 nm).

**Optical measurements.** The optical measurements are performed in an optical cryostat (Quantum Design, OptiCool) with a temperature down to 2 K (nominal). We use a supercontinuum laser (Fianium Femtopower 1060 Supercontinuum Laser) as the light source for the reflection spectroscopy. The laser is focused on the sample by a 20× Mitutoyo objective with ~1.5μm beam size. A small beam size provides a local probe that we can park in a clean region free of bubbles. We choose a very low incident laser power (on the order of 10 nW) to minimize photodoping effects. The spectra are independent of the incident light power up to 200 nW. The reflected light is collected by a spectrometer (Princeton Instruments PIXIS 256e) with 1000 ms exposure time. To minimize the influence of laser fluctuations, another laser beam reflected from a silver mirror is simultaneously collected to normalize the sample reflection spectra.

**Calibration of carrier density.** To build a map between reflection spectrum and carrier density, we perform gate voltage scans at low bias voltage and low vertical electric field. The gap is not closed, so at most one layer is actively doped and the other layer remains intrinsic. In this regime, the gate dependence of the doping density is well understood and can be described by a parallel capacitor model. Taking the hole layer as an example, Extended Data Fig. 5a shows such a scan. When the chemical potential lies in the bandgap (large positive $V_G$), the hole density remains zero, so the spectrum does not change with $V_G$. Upon a critical voltage, the chemical potential is at the valence band edge, and the spectrum starts to change. We first determine this critical voltage from gate dependence of $\| R(\lambda) - R_{\text{intrinsic}}(\lambda) \|$, where $\lambda$ denotes wavelength and $\| \cdots \|$ means vector norm, as shown in Extended Data Fig. 5b. This quantity is a measure of the spectrum change relative to the intrinsic spectrum. We fit the initial increase to a polynomial. The crossing point of this polynomial and the constant baseline gives the critical voltage.

When the gate voltage is further decreased, holes start to fill the valence band. The hole density is then solved from $\mu_h(n_h) + e\phi_h(n_h) = eV_h - E_v$, as shown in Extended Data Fig. 5b. Hartree-Fock theory gives the relation between the chemical potential and density[30]:

$$\frac{d\mu_h}{dn_h} = \frac{\pi \hbar^2}{m_h^*} - \frac{e^2}{4\pi\epsilon_0 \epsilon_h} \sqrt{\frac{2}{\pi n_h}}, \tag{8}$$

where $\hbar$ is the reduced Planck constant, and $m_h^*$ is the hole effective mass. The first term is the non-interacting kinetic energy, and the second term describes the Coulomb interactions. $\epsilon_h = 5.7$ is the effective dielectric constant for the hole layer, which is given by the following procedure: We build a layered dielectric environment according to our device geometry, where the anisotropic dielectric constant of hBN, WSe$_2$ and MoSe$_2$ are all considered[35]. The screening from the graphite gates is included by applying Dirichlet boundary condition for the electrostatic potential. A point charge is placed inside the WSe$_2$ layer, and we numerically solve the Poisson's equation to get the electric potential as a function of the in-plane distance $r$ to the point charge. This potential is fitted to a $1/r$ decay to get the effective dielectric constant.

In fact, the quantum capacitance is much larger than the geometrical capacitance in our device geometry. Even if we ignore this intralayer correlation effect, the carrier density calibration only changes by ~10%. Extended Data Fig. 5c displays the WSe$_2$ reflection spectrum as a function of hole density. The electron density in the MoSe$_2$ layer is determined in the same way (Extended Data Fig. 5d).

**Mean-field theory.** The mean-field theory calculations are performed in the same way as in Reference 6. We self-consistently solve the mean-field Hamiltonian $H_{MF} = H_0 + H_X$ in which

$H_0 = \sum_{\mathbf{k}}(\frac{\hbar^2 k^2}{2m^*} + \frac{E_g}{2})(a^\dagger_{c\mathbf{k}} a_{c\mathbf{k}} - a^\dagger_{v\mathbf{k}} a_{v\mathbf{k}})$, the single-particle contribution, contains parabolically dispersing conduction and valence bands with equal effective masses $m^* = 0.5\, m_e$, and the exchange self-energy $H_X = -\frac{1}{A}\sum_{l',l,\mathbf{k}',\mathbf{k}} V_{l'l}(\mathbf{k}' - \mathbf{k})\delta\rho_{l'l}(\mathbf{k}')a^\dagger_{l'\mathbf{k}} a_{l\mathbf{k}}$ where $a^\dagger, a$ are the creation and annihilation operators, and $l', l = c, v$ are the band (layer) indices. The density-matrix $\delta\rho_{l'l}(\mathbf{k}')$ is the difference between the density matrix of the negative energy band and the density matrix of a full valence and empty conduction band. The gap $E_g$ is a tuning parameter that controls the exciton density. The interaction $V_{l'l}(\mathbf{q})$ takes the dual-gate-screened Coulomb form that distinguishes intralayer and interlayer interactions:

$$V_{l'l}(\mathbf{q}) = \begin{cases} \dfrac{2\pi e^2}{\epsilon q} \dfrac{(e^{qd} - e^{-qD})(e^{-qd} - e^{-qD})}{1 - e^{-2qD}}, & (l' = l), \\ \dfrac{2\pi e^2}{\epsilon q} \dfrac{e^{qd}(e^{-qd} - e^{-qD})^2}{1 - e^{-2qD}}, & (l' \neq l), \end{cases} \quad (8)$$

where $d$ is the distance between the electron and hole layers and $D$ is the distance between the top and back gates. The dielectric constant of hBN has in-plane component $\epsilon_\perp = 7$ and out-of-plane component $\epsilon_{zz} = 4.2$. The quantities $\epsilon, d$, and $D$ that appear in the above equation are replaced by their effective values $\epsilon = \sqrt{\epsilon_\perp \epsilon_{zz}}$, $d = d_0\sqrt{\epsilon_\perp/\epsilon_{zz}}$, and $D = D_0\sqrt{\epsilon_\perp/\epsilon_{zz}}$ where $d_0$ and $D_0$ are the physical distances. In our calculations we take $d_0 = 3$ nm and $D_0 = 40$ nm. The density matrix $\rho$ is defined relative to the state in which the valence band is filled and conduction band empty: $\rho_{l'l}(\mathbf{k}) = \langle a^\dagger_{l\mathbf{k}} a_{l'\mathbf{k}} \rangle - \delta_{l'l}\delta_{lv}$.

The electron and hole chemical potentials are defined relative to the conduction band bottom and valence band top respectively: $\mu_e = \mu - \frac{E_g}{2}$, $\mu_h = -\mu - \frac{E_g}{2}$ where $\mu$ is the chemical potential in the mean-field calculations. The electrostatic (Hartree) potential is not included in the definition of chemical potentials because it has already been accounted for in the gate-geometry dependent electric potential calculation. This separation of the chemical potential into a gate-geometry dependent part and a part that depends only on the interacting electrons and holes is necessitated by the long-range of the Coulomb interaction.

## References

1. Zhang, Y., Tan, Y.-W., Stormer, H. L. & Kim, P. Experimental observation of the quantum Hall effect and Berry's phase in graphene. *Nature* **438**, 201–204 (2005).
2. Klitzing, K. v., Dorda, G. & Pepper, M. New Method for High-Accuracy Determination of the Fine-Structure Constant Based on Quantized Hall Resistance. *Phys. Rev. Lett.* **45**, 494–497 (1980).
3. Andrei, E. Y. *et al.* Observation of a Magnetically Induced Wigner Solid. *Phys. Rev. Lett.* **60**, 2765–2768 (1988).
4. Smoleński, T. *et al.* Signatures of Wigner crystal of electrons in a monolayer semiconductor. *Nature* **595**, 53–57 (2021).
5. Zhou, Y. *et al.* Bilayer Wigner crystals in a transition metal dichalcogenide heterostructure. *Nature* **595**, 48–52 (2021).
6. Zeng, Y. & MacDonald, A. H. Electrically controlled two-dimensional electron-hole fluids. *Phys. Rev. B* **102**, 085154 (2020).

7. Jauregui, L. A. *et al.* Electrical control of interlayer exciton dynamics in atomically thin heterostructures. *Science* **366**, 870–875 (2019).
8. Paik, E. Y. *et al.* Interlayer exciton laser of extended spatial coherence in atomically thin heterostructures. *Nature* **576**, 80–84 (2019).
9. Rivera, P. *et al.* Observation of long-lived interlayer excitons in monolayer MoSe2–WSe2 heterostructures. *Nat. Commun.* **6**, 6242 (2015).
10. Conti, S. *et al.* Chester Supersolid of Spatially Indirect Excitons in Double-Layer Semiconductor Heterostructures. *Phys. Rev. Lett.* **130**, 057001 (2023).
11. Lozovik, Y. E., Kurbakov, I. L., Astrakharchik, G. E., Boronat, J. & Willander, M. Strong correlation effects in 2D Bose–Einstein condensed dipolar excitons. *Solid State Commun.* **144**, 399–404 (2007).
12. Wu, F.-C., Xue, F. & MacDonald, A. H. Theory of two-dimensional spatially indirect equilibrium exciton condensates. *Phys. Rev. B* **92**, 165121 (2015).
13. Perali, A., Neilson, D. & Hamilton, A. R. High-Temperature Superfluidity in Double-Bilayer Graphene. *Phys. Rev. Lett.* **110**, 146803 (2013).
14. Ma, L. *et al.* Strongly correlated excitonic insulator in atomic double layers. *Nature* **598**, 585–589 (2021).
15. Zhang, Z. *et al.* Correlated interlayer exciton insulator in heterostructures of monolayer WSe2 and moiré WS2/WSe2. *Nat. Phys.* 1–7 (2022) doi:10.1038/s41567-022-01702-z.
16. Astrakharchik, G. E., Boronat, J., Kurbakov, I. L. & Lozovik, Yu. E. Quantum Phase Transition in a Two-Dimensional System of Dipoles. *Phys. Rev. Lett.* **98**, 060405 (2007).
17. Wang, G. *et al.* Colloquium: Excitons in atomically thin transition metal dichalcogenides. *Rev. Mod. Phys.* **90**, 021001 (2018).
18. Regan, E. C. *et al.* Emerging exciton physics in transition metal dichalcogenide heterobilayers. *Nat. Rev. Mater.* **7**, 778–795 (2022).
19. Chernikov, A. *et al.* Exciton binding energy and nonhydrogenic Rydberg series in monolayer $WS_2$. *Phys. Rev. Lett.* **113**, 076802 (2014).
20. Calman, E. V. *et al.* Indirect excitons in van der Waals heterostructures at room temperature. *Nat. Commun.* **9**, 1895 (2018).
21. Fogler, M. M., Butov, L. V. & Novoselov, K. S. High-temperature superfluidity with indirect excitons in van der Waals heterostructures. *Nat. Commun.* **5**, 4555 (2014).
22. Debnath, B., Barlas, Y., Wickramaratne, D., Neupane, M. R. & Lake, R. K. Exciton condensate in bilayer transition metal dichalcogenides: Strong coupling regime. *Phys. Rev. B* **96**, 174504 (2017).
23. Mak, K. F. *et al.* Tightly bound trions in monolayer MoS2. *Nat. Mater.* **12**, 207–211 (2012).
24. Scuri, G. *et al.* Large Excitonic Reflectivity of Monolayer ${\mathrm{MoSe}}_{2}$ Encapsulated in Hexagonal Boron Nitride. *Phys. Rev. Lett.* **120**, 037402 (2018).
25. Ross, J. S. *et al.* Electrical control of neutral and charged excitons in a monolayer semiconductor. *Nat. Commun.* **4**, 1474 (2013).
26. Christopher, J. W., Goldberg, B. B. & Swan, A. K. Long tailed trions in monolayer MoS2: Temperature dependent asymmetry and resulting red-shift of trion photoluminescence spectra. *Sci. Rep.* **7**, 14062 (2017).

27. Zhang, C., Wang, H., Chan, W., Manolatou, C. & Rana, F. Absorption of light by excitons and trions in monolayers of metal dichalcogenide Mo S 2 : Experiments and theory. *Phys. Rev. B* **89**, 205436 (2014).
28. Liu, E. *et al.* Exciton-polaron Rydberg states in monolayer MoSe2 and WSe2. *Nat. Commun.* **12**, 6131 (2021).
29. Goryca, M. *et al.* Revealing exciton masses and dielectric properties of monolayer semiconductors with high magnetic fields. *Nat. Commun.* **10**, 4172 (2019).
30. Nagano, S., Singwi, K. S. & Ohnishi, S. Correlations in a two-dimensional quantum electron gas: The ladder approximation. *Phys. Rev. B* **29**, 1209–1213 (1984).
31. Li, L. *et al.* Very Large Capacitance Enhancement in a Two-Dimensional Electron System. *Science* **332**, 825–828 (2011).
32. Kravchenko, S. V., Rinberg, D. A., Semenchinsky, S. G. & Pudalov, V. M. Evidence for the influence of electron-electron interaction on the chemical potential of the two-dimensional electron gas. *Phys. Rev. B* **42**, 3741–3744 (1990).
33. Regan, E. C. *et al.* Mott and generalized Wigner crystal states in WSe2/WS2 moiré superlattices. *Nature* **579**, 359–363 (2020).
34. Kappei, L., Szczytko, J., Morier-Genoud, F. & Deveaud, B. Direct Observation of the Mott Transition in an Optically Excited Semiconductor Quantum Well. *Phys. Rev. Lett.* **94**, 147403 (2005).
35. Laturia, A., Van de Put, M. L. & Vandenberghe, W. G. Dielectric properties of hexagonal boron nitride and transition metal dichalcogenides: from monolayer to bulk. *Npj 2D Mater. Appl.* **2**, 1–7 (2018).

## Data availability
The data that support the findings of this study are available from the corresponding author upon reasonable request.

## Acknowledgements
This work was supported primarily by the U.S. Department of Energy, Office of Science, Office of Basic Energy Sciences, Materials Sciences and Engineering Division under contract no. DE-AC02-05-CH11231 (van der Waals heterostructures programme, KCWF16). The data analysis was also supported by the AFOSR award FA9550-23-1-0246. . S.T. acknowledges support from DOE-SC0020653, NSF CMMI 1933214, NSF mid-scale 1935994, NSF 1904716, NSF DMR 1552220 and DMR 1955889. K.W. and T.T. acknowledge support from JSPS KAKENHI (Grant Numbers 19H05790, 20H00354).

## Author contributions
F.W. conceived the research. R.Q. and A.Y.J. fabricated the devices. A.Y.J. and R.Q. performed the optical measurements with help from Z.Z., E.R., J.X, and Z.L.. R.Q., A.Y.J. and F.W. analyzed the data. Y.Z. and A.H.M. did the mean-field calculations. T.Z., Q.F. and M.F.C. contributed to the device fabrication. S.T. grew WSe$_2$ and MoSe$_2$ crystals. K.W. and T.T. grew hBN crystals. All authors discussed the results and wrote the manuscript.

## Competing interests
The authors declare no competing interests.

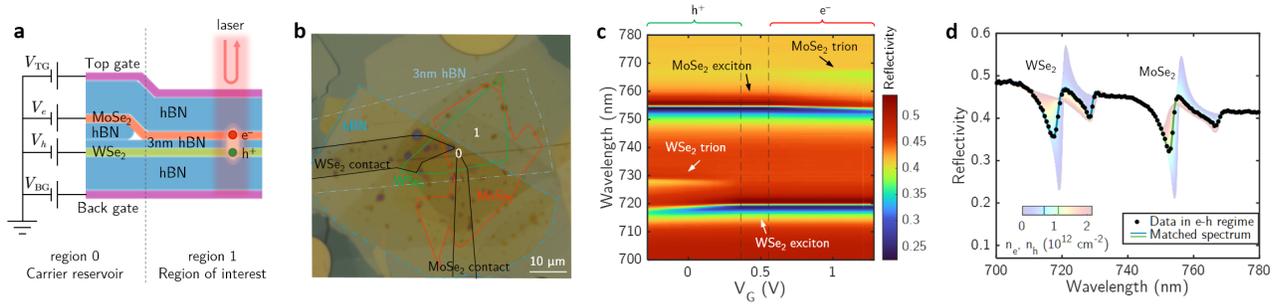

**Fig. 1 | Electrical control and optical detection of charge states.**
**a**, Schematic cross-section of the MoSe$_2$/hBN/WSe$_2$ heterostructure device.
**b**, Optical microscopy image of the device with layer boundaries outlined.
**c**, Typical gate-dependent reflection spectra at a low bias voltage ($V_\mathrm{B} = 0.75$ V).
**d**, Determination of carrier densities for a typical reflection spectrum in the e-h fluid regime. Black dots, reflection spectrum at $V_\mathrm{B} = 0.87$ V, $V_\mathrm{G} = 0.92$ V. Faded colored lines, spectrum-density maps for WSe$_2$ (left) and MoSe$_2$ (right) determined from single-layer doped regime. Solid lines, reflection spectrum from the spectra-density maps that achieves the best match with the black dots.

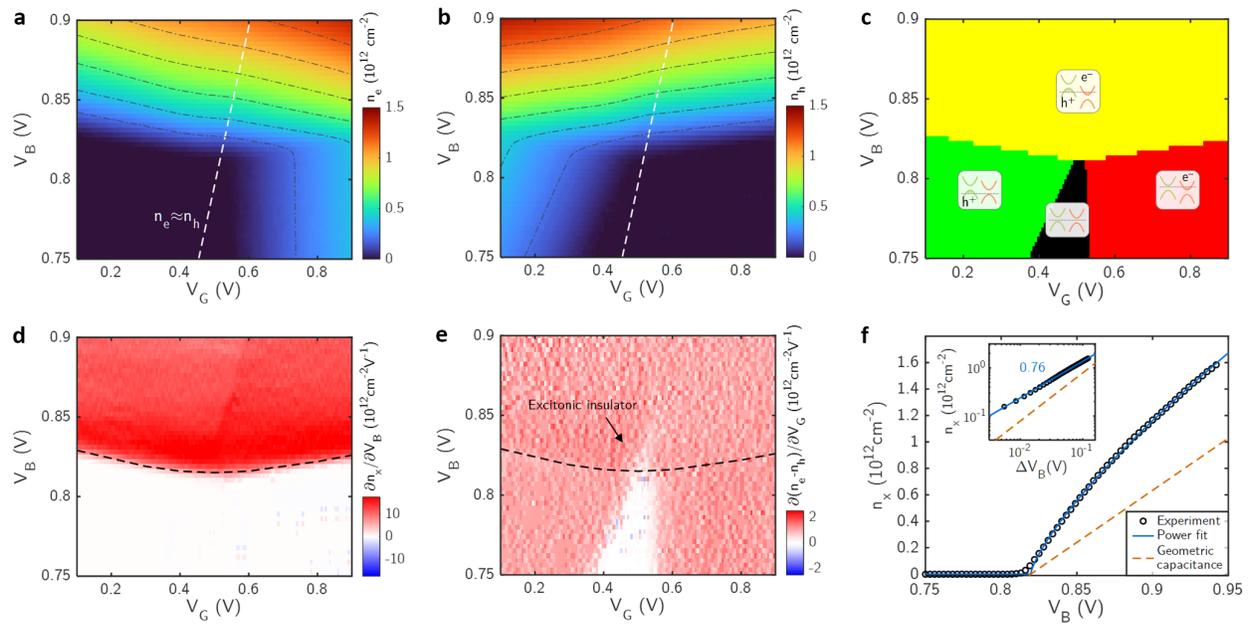

**Fig. 2 | Carrier density and compressibility maps.**

**a-b**, False color map of experimentally determined electron (**a**) and hole (**b**) densities. Contour lines are overlayed as gray dash-dotted curves. The dashed white line is along $n_e \approx n_h$.

**c**, Phase diagram of the e-h system determined experimentally using the density of $0.002 \times 10^{12}$ cm$^{-2}$ as a threshold. Black: no carrier present in the system. Green: WSe$_2$ is hole doped. Red: MoSe$_2$ is electron doped. Yellow: both electrons and holes are present. Four insets schematically show the band alignment in each phase.

**d**, Partial derivative of the exciton density $n_x = \min(n_e, n_h)$ with respect to bias voltage $V_B$, corresponding to the interlayer exciton compressibility.

**e**, Partial derivative of net charge density $n_e - n_h$ with respect to gate voltage $V_G$, corresponding to the charge compressibility.

**f**, Bias dependence of exciton density along $n_e \approx n_h$ line. The blue curve is a power function fit to the experimental data (scatters). Yellow dashed line, density determined from the geometric capacitance, showing a linear increase with the bias voltage. Inset: same data but plotted in log-log scale, showing a power-law scaling.

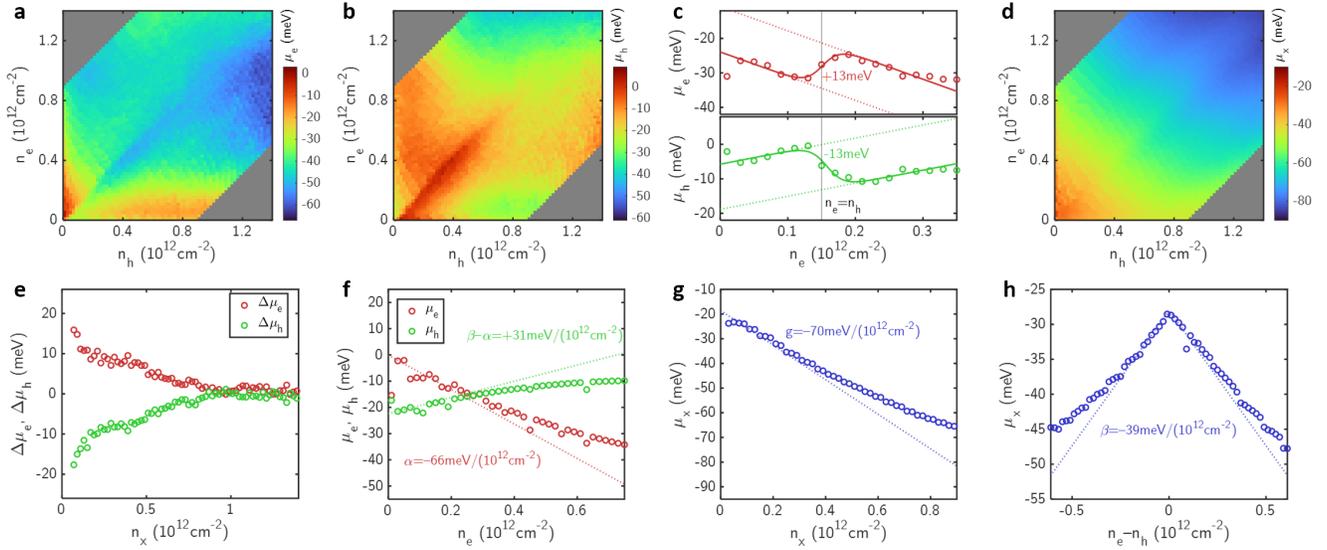

**Fig. 3 | Chemical potential of the correlated e-h fluids.**

**a-b**, Measured chemical potential maps for electrons (**a**) and holes (**b**).

**c**, Line cut of the electron and hole chemical potentials at fixed hole density $n_h = 0.15 \times 10^{12}$ cm$^{-2}$. The experimental data (empty circles) is fitted to a linear background plus a sigmoid function (solid lines).

**d**, Measured exciton chemical potential map $\mu_x = \mu_e + \mu_h$.

**e**, Chemical potential jump at net charge neutrality for different exciton densities.

**f**, Electron chemical potential $\mu_e(n_e, n_h = 0)$ and hole chemical potential $\mu_h(n_e, n_h = 0)$ as a function of electron density, keeping the hole density zero.

**g**, Exciton chemical potential as a function of exciton density, keeping electron and hole densities equal.

**h**, Exciton chemical potential as a function of unpaired charge density $n_e - n_h$, keeping exciton density $n_x = 0.05 \times 10^{12}$ cm$^{-2}$ constant. Dotted lines in **f-h** are linear expansions of the experimental data in the low doping region ($0.05 \times 10^{12}$ cm$^{-2} \leq n \leq 0.5 \times 10^{12}$ cm$^{-2}$).

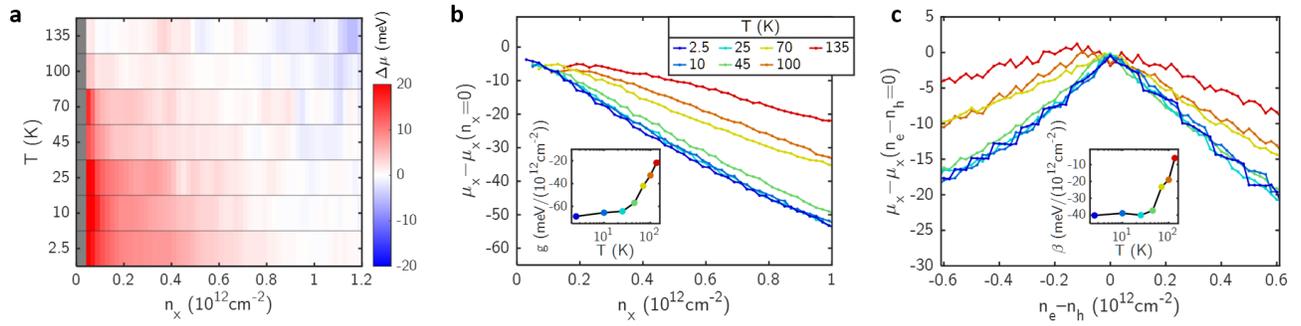

**Fig. 4 | Temperature dependence.**

**a**, Chemical potential jump at net charge neutrality as a function of exciton density and temperature.

**b**, Exciton density dependence of the exciton chemical potential at various temperatures. The electron and hole densities are kept equal. Inset: Fitted exciton-exciton interaction strength $g$ as a function of temperature.

**c**, Exciton chemical potential as a function of unpaired charge density $n_e - n_h$ at constant exciton density $n_x = 0.05 \times 10^{12}$ cm$^{-2}$. Inset: Fitted exciton-charge interaction strength $\beta$ as a function of temperature.

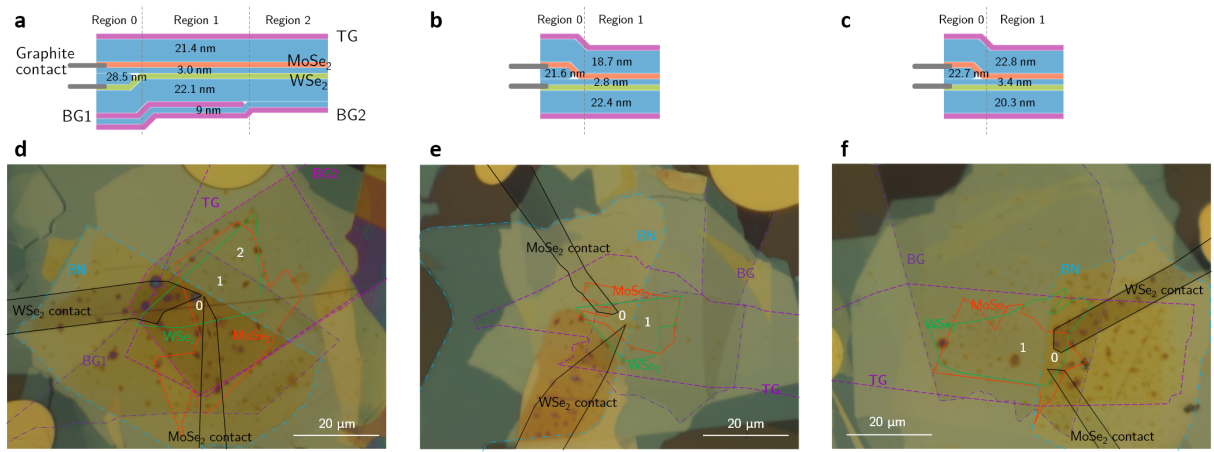

**Extended Data Fig. 1 | Device structure and optical images.**
**a-c**, Schematic cross section of devices D1 (**a**), D2 (**b**) and D3 (**c**).
**d-f**, Optical microscope images of the devices, with flake boundaries outlined.

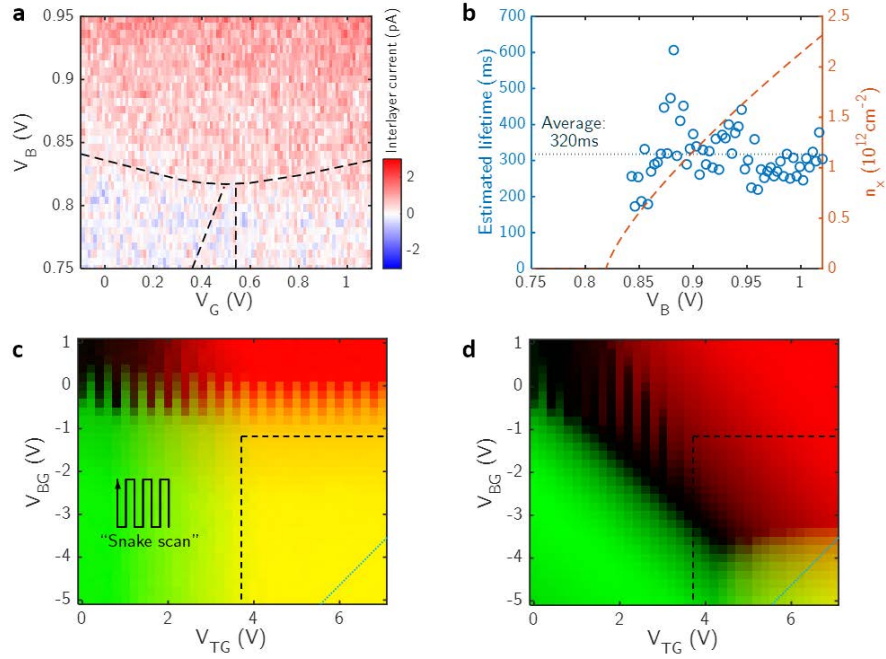

**Extended Data Fig. 2 | Equilibrium electron-hole fluids.**

**a**, Interlayer leakage current as a function of the bias and gate voltage.

**b**, Left axis: Estimated recombination lifetime calculated from the interlayer leakage current along $n_e \approx n_h$ line. Right axis: Interlayer exciton density along the same line cut.

**c-d**, Region 0 and region 1 doping phase diagram at constant $V_B = 1$ V. Red and green channels of the image indicate the electron and hole doping, respectively. The data is taken with alternating scan direction for neighboring columns (snake scan). Inside the dashed boxes, region 0 is at high doping for both layers, and region 1 shows no hysteresis. Blue dotted line indicates where the data in the main text is taken.

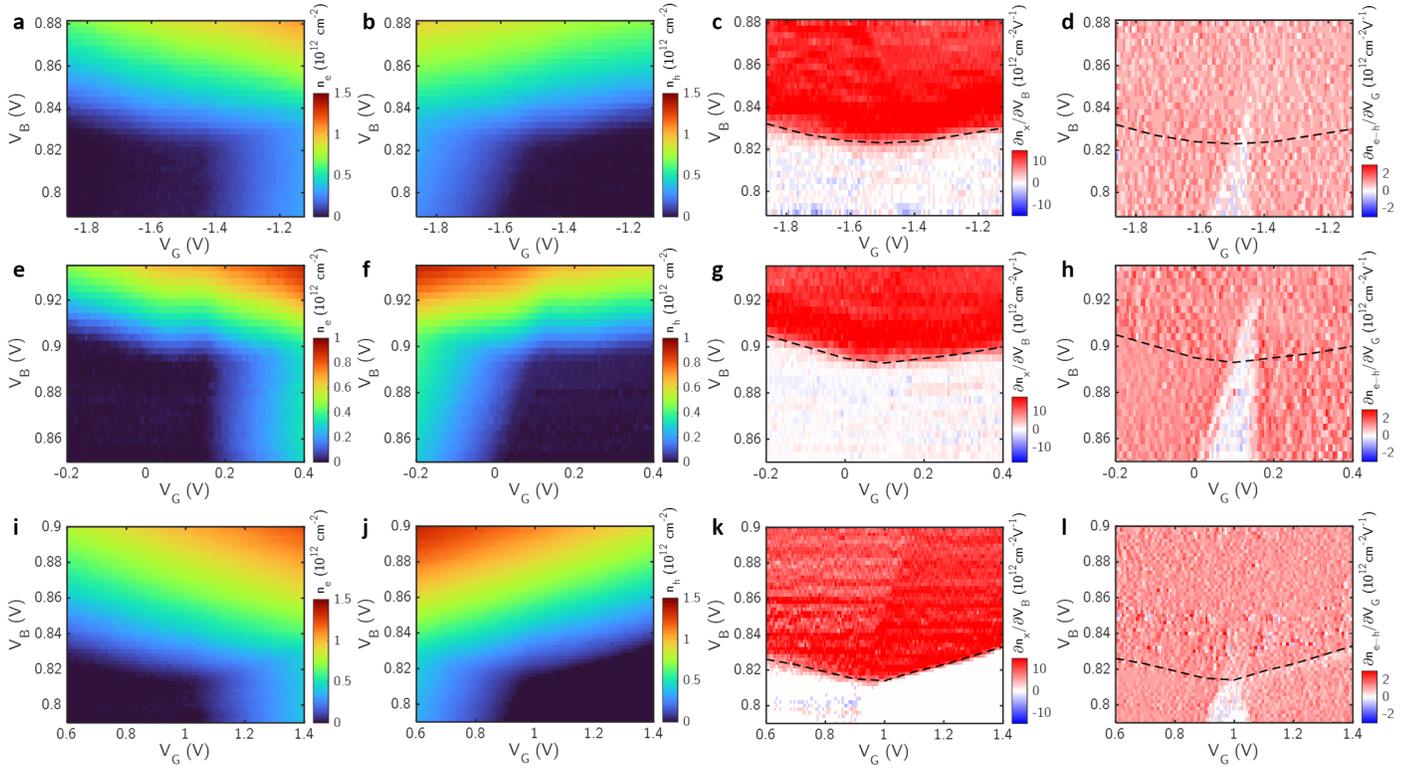

**Extended Data Fig. 3 | Density maps from other devices.**

**a-d**, Measured electron density **(a)**, hole density **(b)**, exciton compressibility **(c)**, and charge compressibility **(d)** from D1 region 2 with $V_\Delta = 13$ V.

**e-h**, Same data from D2 with $V_\Delta = 9$ V.

**i-l**, Same data from D3 with $V_\Delta = 9$ V. This device has slightly thicker hBN layer, and the exciton insulator phase becomes less obvious.

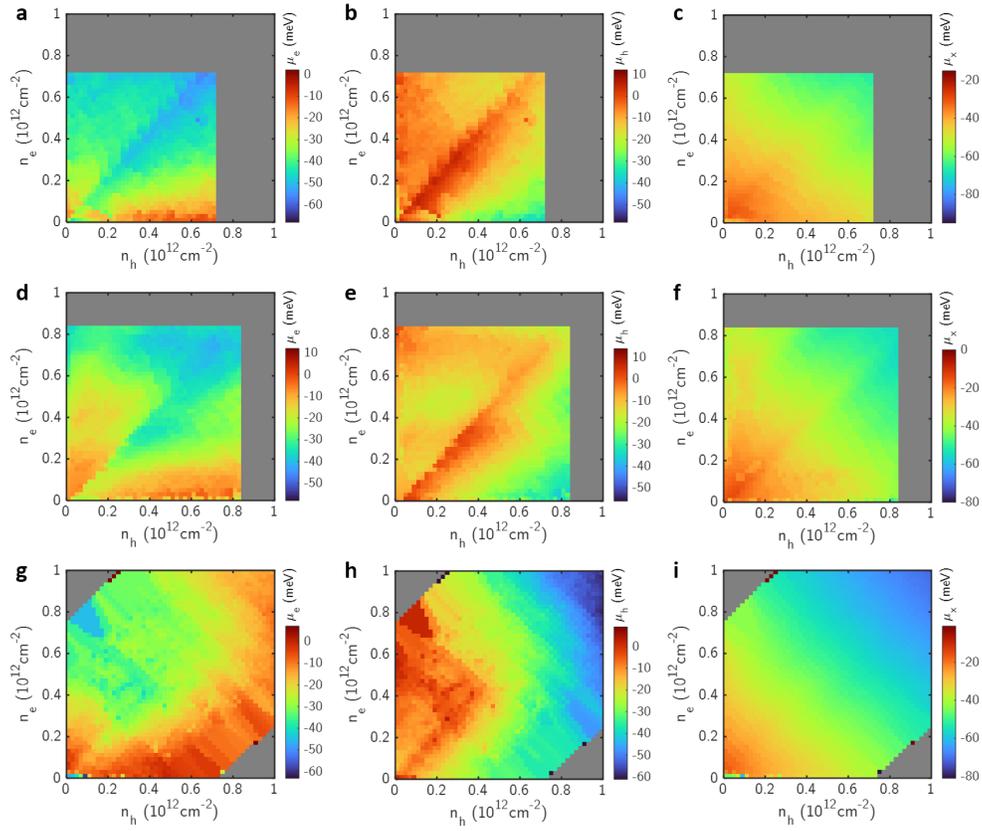

**Extended Data Fig. 4 | Experimental chemical potential from other devices.**
**a-c**, Electron **(a)**, hole **(b)**, and exciton **(c)** chemical potentials extracted from D1 region 2.
**d-f**, Same data from D2.
**g-i**, Same data from D3. The chemical potential jump becomes weaker with thicker hBN spacer.

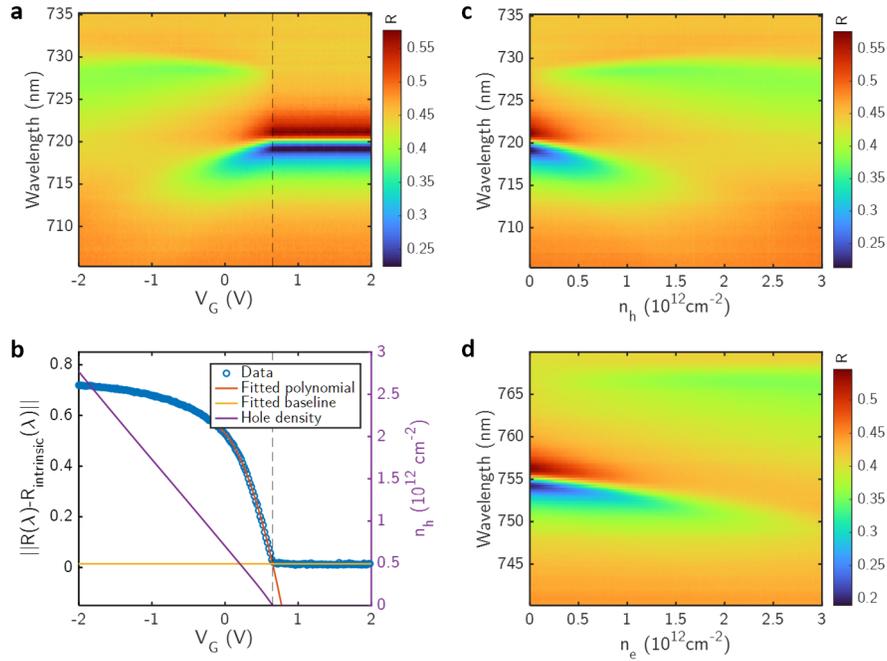

**Extended Data Fig. 5 | Carrier density calibration.**

**a**, Gate dependence of the absorption spectrum at $V_B = 1\,\text{V}$, $V_\Delta = 7\,\text{V}$. The type-II band gap is not closed under this condition, so at most one layer is doped.

**b**, Determination of the critical voltage that the chemical potential is on the valence band edge. Blue scatters, change of reflection spectrum relative to charge neutrality. Yellow and red curves are fitted baseline and fitted polynomial curves that capture the initial change of the spectrum after the start of the doping. The vertical dashed line is the determined voltage at which the doping begins. Purple line is the calibrated hole density based on the model described in Methods section.

**c-d**, WSe$_2$ and MoSe$_2$ reflection spectra as a function of the charge density.

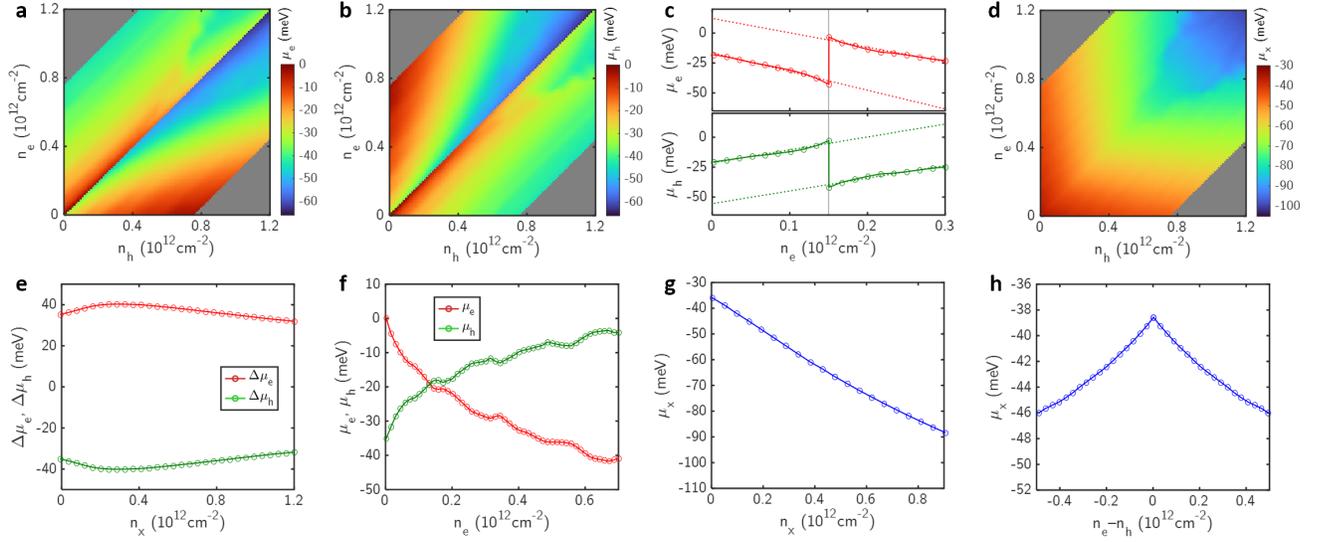

**Extended Data Fig. 6 | Mean-field theory calculation.**

**a-b**, Calculated chemical potential maps for electrons (**a**) and holes (**b**).

**c**, Line cut of the electron and hole chemical potentials at fixed hole density $n_h = 0.15 \times 10^{12}$ cm$^{-2}$.

**d**, Calculated exciton chemical potential map $\mu_x = \mu_e + \mu_h$.

**e**, Chemical potential jump at net charge neutrality for different exciton densities. The mean-field theory does not capture the Mott transition due to lack of consideration of screening.

**f**, Electron chemical potential $\mu_e(n_e, n_h = 0)$ and hole chemical potential $\mu_h(n_e, n_h = 0)$ as a function of electron density, keeping the hole density zero.

**g**, Exciton chemical potential as a function of exciton density, keeping electron and hole densities equal.

**h**, Exciton chemical potential as a function of unpaired charge density $n_e - n_h$, keeping exciton density $n_x = 0.05 \times 10^{12}$ cm$^{-2}$ constant.

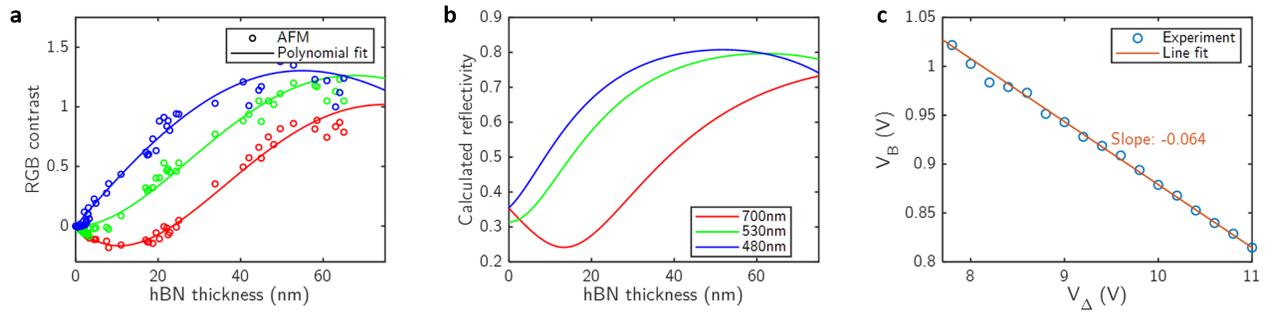

**Extended Data Fig. 7 | hBN thickness calibration.**

**a**, Scatter plot of the measured red, green, and blue (RGB) channels of the hBN optical contrast for different measured hBN thicknesses. The solid lines are 4$^{th}$ order polynomial fits to the experimental points.

**b**, Calculated reflectivity based on Fresnel equations. The system consists of (from top to bottom) semi-infinite air, hBN with variable thickness, 90 nm SiO$_2$ and semi-infinite Si. Normal incidence is assumed.

**c**, Bias voltage required to close the band gap as a function of $V_\Delta$. The slope gives the negative ratio of thin hBN thickness to the total hBN thickness (see Methods for details).